\documentclass{webofc}
\usepackage[varg]{txfonts}   
\usepackage{comment}
\usepackage{nicefrac}
\begin{document}

\title{Heavy quark momentum diffusion coefficient during hydrodynamization via effective kinetic theory}

\author{ 
\firstname{Kirill} \lastname{Boguslavski}\inst{1}
\and
     \firstname{Aleksi} \lastname{Kurkela}\inst{2}
     \and
            \firstname{Tuomas} \lastname{Lappi}\inst{3,4}
      \and
 \firstname{Florian} \lastname{Lindenbauer}\inst{1}
 \and 
\firstname{Jarkko} \lastname{Peuron}\inst{3,4}\fnsep\thanks{\email{jarkko.t.peuron@jyu.fi}}
}

\institute{
Institute for Theoretical Physics, Technische Universit\"{a}t Wien, 1040 Vienna, Austria
\and
Faculty of Science and Technology, University of Stavanger, 4036 Stavanger, Norway
\and
Department of Physics, P.O.~Box 35, 40014 University of Jyv\"{a}skyl\"{a}, Finland
\and
           Helsinki Institute of Physics, P.O.~Box 64, 00014 University of Helsinki, Finland
          }

\newcommand{\xx}{{\mathbf{x}}}
\newcommand{\pp}{{\mathbf{p}}}
\newcommand{\qq}{{\mathbf{q}}}
\newcommand{\ptt}{{p_T}}
\newcommand{\ktt}{{k_T}}
\newcommand{\ud}{\mathrm{d}}
\newcommand{\uc}{{\mathrm{c}}}
\newcommand{\ul}{{\mathrm{L}}}
\newcommand{\intd}{\int \!}
\newcommand{\tr}{\, \mathrm{Tr} \, }
\newcommand{\R}{\mathrm{Re}}
\newcommand{\nc}{{N_\mathrm{c}}}
\newcommand{\nf}{{N_\mathrm{F}}}
\newcommand{\half}{\frac{1}{2}}
\newcommand{\ah}{\mathrm{ah}}
\newcommand{\hc}{\mathrm{\ h.c.\ }}
\newcommand{\hctr}{\mathrm{\ h.c.;tr \ }}
\newcommand{\nosum}[1]{\textrm{ (no sum over } #1 )}
\newcommand{\na}{\, :\!}
\newcommand{\nb}{\!: \,}
\newcommand{\cf}{C_\mathrm{F}}
\newcommand{\ca}{C_\mathrm{A}}
\newcommand{\df}{d_\mathrm{F}}
\newcommand{\da}{d_\mathrm{A}}
\newcommand{\bs}[1]{\boldsymbol{#1}}
\newcommand{\qs}{Q_{\mathrm{s}}}
\newcommand{\as}{\alpha_{\mathrm{s}}}
\newcommand{\lqcd}{\Lambda_{\mathrm{QCD}}}
\newcommand{\nr}[1]{(\ref{#1})} 
\newcommand*{\rom}[1]{\expandafter\@slowromancap\romannumeral #1@}
\newcommand{\mbf}{\mathbf}
\newcommand{\tout}{t_{\text{out}}}
\newcommand{\tpert}{t_{\text{pert}}}
\newcommand{\tcent}{\bar{t}}			
\newcommand{\mrm}{\mathrm}
\newcommand{\Tr}{\mrm{Tr}}
\newcommand{\HTL}{\mrm{HTL}}
\newcommand{\fit}{\mrm{fit}}
\newcommand{\rel}{\mrm{rel}}
\newcommand{\pInit}{p_0}
\newcommand{\Q}{Q}
\newcommand{\wplas}{\omega_{\mrm{pl}}}
\newcommand{\fig}{Fig.~}
\newcommand{\figs}{Figs.~}
\newcommand{\eq}{Eq.~}
\newcommand{\se}{Sec.~}
\newcommand{\eqs}{Eqs.~}
\newcommand{\re}{Ref.~}
\newcommand{\res}{Refs.~}
\newcommand{\ii}{{\boldsymbol{\hat{\i}}}}
\newcommand{\jj}{{\boldsymbol{\hat{\j}}}}
\newcommand{\kk}{{\mathbf{\hat{k}}}}
\newcommand{\cl}{\mrm{cl}}
\newcommand{\uj}{\mrm{j}}
\newcommand{\dtmax}{\Delta t_{\text{max}}}
\newcommand{\tstar}{T_*}
\newcommand{\der}{\mathrm{d}}
\newcommand{\pmin}{p_{\mathrm{min}}}
\newcommand{\nn}{\nonumber}
\newcommand{\qperp}{q_\perp}
\newcommand{\vqperp}{\vec q_\perp}
\newcommand{\lperp}{\Lambda_\perp} 
\newcommand{\qhat}{\hat q}
\newcommand{\qhatf}{\qhat_{\mathrm{f}}}
\newcommand{\qhatff}{\qhat_{\mathrm{ff}}}
\newcommand{\kmin}{k_{\mathrm{min}}}
\newcommand{\Gammael}{\Gamma_{\text{el}}}
\newcommand{\vecp}{\vec p}
\newcommand{\vb}{\vec}
\renewcommand{\vec}[1]{\mathrm{\mathbf{#1}}}
\newcommand{\dd}[2][]{\mathrm d^{#1}{#2}\,} 
\newcommand{\dv}[2][]{\frac{\dd{#1}}{\dd{#2}}}
\newcommand{\pdv}[2][]{\frac{\partial{#1}}{\partial{#2}}}
\newcommand{\pmax}{p_{\mathrm{max}}}
\newcommand{\Ejet}{E_{\mathrm{jet}}}
\newcommand{\taubmss}{\tau_{\mathrm{BMSS}}}
\newcommand{\tauR}{\tau_R}
\newcommand{\Conetwo}{\mathcal {C}^{1\leftrightarrow 2}}
\newcommand{\Ctwotwo}{\mathcal{ C}^{2\leftrightarrow 2}}
\newcommand{\Cexp}{\mathcal C^{\mathrm{exp}}}
\newcommand{\Teps}{T_{\varepsilon}}
\newcommand{\tauT}{\tau_{\mathrm{BMSS}}}

\abstract{%
  In these proceedings, we compute the heavy quark momentum diffusion coefficient 
 using QCD effective kinetic theory for a plasma going through the bottom-up thermalization scenario until approximate hydrodynamization. This transport coefficient describes heavy quark momentum diffusion in the quark-gluon plasma and is used in many phenomenological frameworks, e.g. in the open quantum systems approach. Our extracted nonthermal diffusion coefficient matches the thermal one for the same energy density within 30\%. At large occupation numbers in the earliest stage, the transverse diffusion coefficient dominates, while the longitudinal diffusion coefficient is larger for the underoccupied system in the later stage of hydrodynamization.
}
\maketitle
\section{Introduction}
\label{intro}
Recent studies on the very early stages of ultrarelativistic heavy-ion collisions indicate that transport coefficients are large during the glasma stage \cite{Carrington:2022bnv,Carrington:2021dvw,Boguslavski:2020tqz,Ipp:2020nfu,Khowal:2021zoo,Sun:2019fud,Boguslavski:2023alu}, where the heavy quark diffusion coefficient is estimated to be roughly $\kappa \approx \mathcal{O}(10) \nicefrac{\mathrm{Gev}^2}{\mathrm{fm}}$  \cite{Avramescu:2023qvv}. In equilibrium we expect $\kappa \approx \mathcal{O}(0.1) \nicefrac{\mathrm{Gev}^2}{\mathrm{fm}}$ \cite{Brambilla:2022xbd}, as illustrated in the left panel of \fig \ref{fig:goodpics}.  The discrepancy is due to the  larger energy density at the early stages. The aim of this work \cite{Boguslavski:2023fdm} is to study $\kappa$ during the hydrodynamization process to answer the questions: How large and how anisotropic is $\kappa$?

\def\twofigsize{0.34}
\begin{figure}
\centering
\includegraphics[scale=\twofigsize]{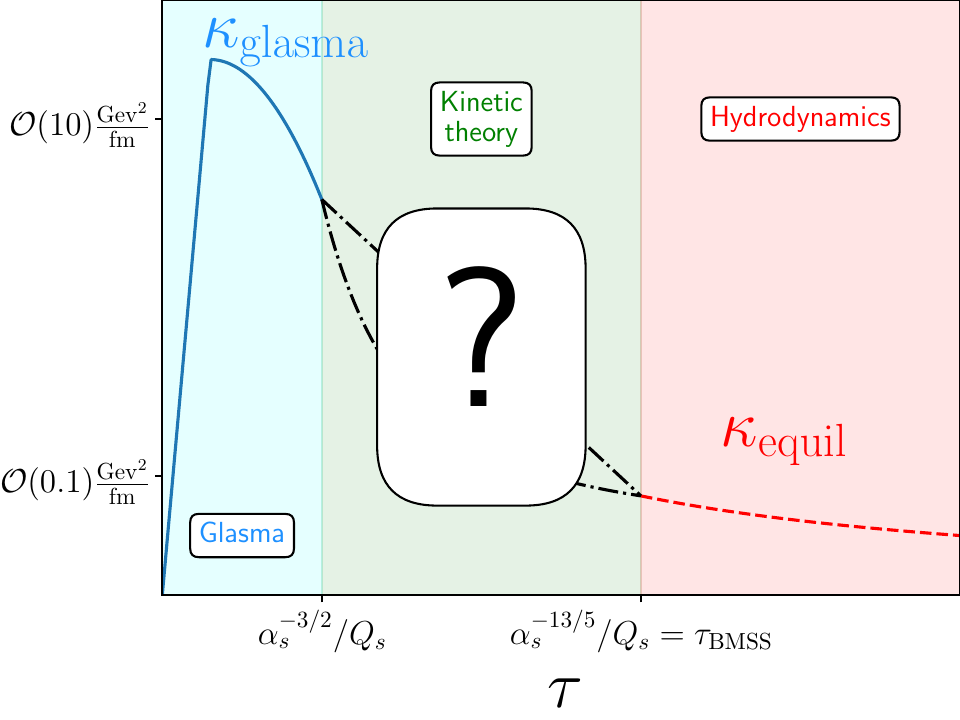}
\includegraphics[scale=\twofigsize]{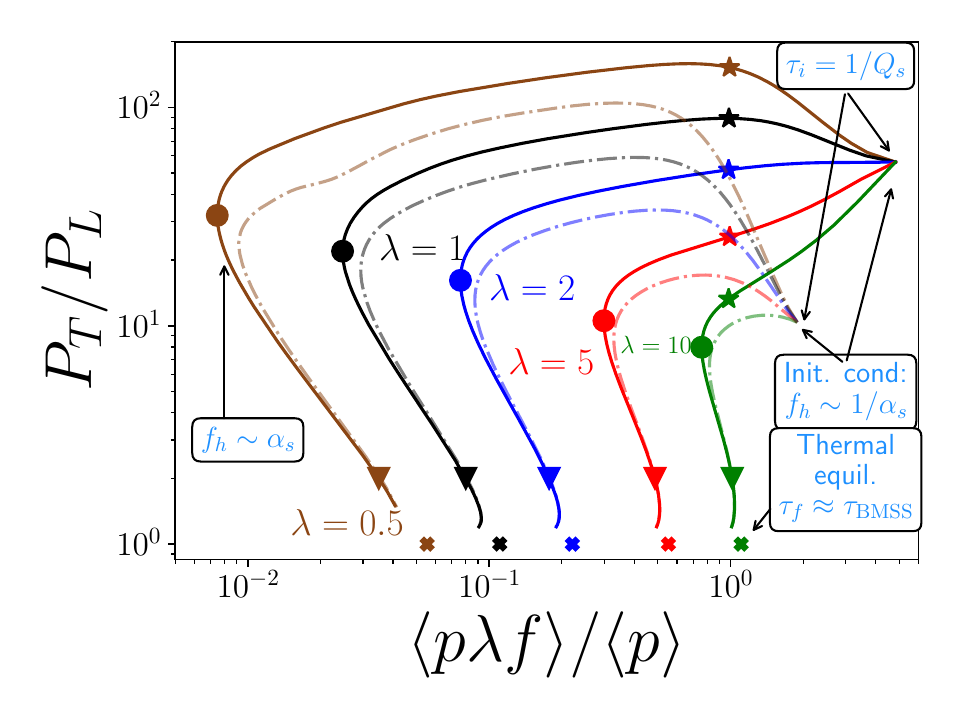}
\caption{Left: Cartoon of the evolution of $\kappa$ from the initial nonequilibrium phase to thermal equilibrium. Right: Trajectory of the system on an occupancy-anisotropy plane \cite{Kurkela:2015qoa}. }
\label{fig:goodpics}
\end{figure}

\section{Theoretical background }
\label{sec:theory}
In effective kinetic theory  \cite{Arnold:2002zm} we describe the evolution of the gluon phase space density $f$ by numerically solving the Boltzmann equation \cite{Kurkela:2015qoa}
\begin{align}
\dfrac{\partial f(\bs{p})}{\partial \tau } = \mathcal{C}_{1 \leftrightarrow 2  }[f]  + \mathcal{C}_{2 \leftrightarrow 2  }[f] + \mathcal{C}_{\mathrm{exp} }[f],
\end{align}
where $\mathcal{C}_{1 \leftrightarrow 2  }$ describes the effective one to two splittings, $\mathcal{C}_{2 \leftrightarrow 2  }$ two to two processes and  $\mathcal{C}_{\mathrm{exp} }$ incorporates the longitudinal expansion in the form of an effective scattering term.
The right panel of \fig \ref{fig:goodpics} illustrates the evolution of the distribution function (different linestyles correspond to different initial conditions) on the occupancy-anisotropy plane. In order to make a connection to the bottom-up thermalization picture \cite{Baier:2000sb}, we use time markers. The star symbol indicates  occupancy $ f_h \sim 1/\lambda$, $\lambda= g^2 N_c $, coinciding with maximum anisotropy for small couplings. The circle marker indicates minimum occupancy.  The triangle marker is located at approximate isotropy, quantified by $P_T/P_L = 2$.

\subsection{Heavy quark diffusion coefficient}
\label{sec:quark}
In kinetic theory, the diffusion coefficient can be computed as \cite{Moore:2004tg}
\begin{align}
\label{eq:KT_omGen}
3\kappa& =  \frac{1}{2M}\int_{\bs{k} \bs{k^\prime} \bs{p^\prime}}\left(2 \pi \right)^3 \delta^3\left( \bs{p} +\bs{k} - \bs{p^\prime} - \bs{k^\prime} \right)  2 \pi \delta \left(k^\prime - k \right) \bs{q}^2 
\left[ \left| \mathcal{M}_\kappa \right|^2 f(\bs{k}) (1+f(\bs{k^\prime})) \right].
\end{align}
Here $q$ is the momentum transfer, $k$ and $k^\prime$ ($p$ and $p^\prime$) are the in- and outgoing gluon (heavy quark) momenta.  We use the shorthand notation $\int_{\bs{p}} =  \int \nicefrac{\der p^3}{2 p^0\left(2 \pi \right)^3}.$  The dominant contributions in the limit of very large quark mass $M$ arise from t-channel gluon exchange and are described by the matrix element 
$
 \left|\mathcal{M}_{\mathrm{\kappa}} \right|^2 = \left[N_c C_H g^4 \right] \dfrac{16 M^2 k^2 \left( 1 + \cos^2 \theta_{\bs{k}\bs{k}'} \right)}{(q^2+m_D^2)^2}.
$
In this limit, the momentum transfer is purely spatial and the screening can be implemented by inserting the screening mass $m_D$ into the propagator. 
The transverse $\kappa_T$ and longitudinal $\kappa_z$ coefficients are related to the full coefficient by 
$
3 \kappa = 2 \kappa_T + \kappa_z.
$

In order to better understand the nonequilibrium medium, we define three scales associated to it. The effective temperature is given by 
$
\tstar = \nicefrac{4\lambda}{m_D}  \int_{\bs{p}} p f(p) (1+f(p)).
$
The Debye screening mass can be computed as $
m_D^2 = 8 \int_{\bs{p}} \lambda f(p).
$
The temperature can be defined from the energy density by
$ T_\varepsilon = \left(\nicefrac{30\,\varepsilon}{\pi^2\nu_g} \right)^{1/4},$
where $\nu_g = 2 \left(\nc^2 -1\right)$ for pure glue QCD.

\section{Results}
\subsection{Comparing non-equilibrium $\kappa$ results to thermal equilibrium}
We compare our nonequilibrium simulations with thermal systems for the same $ \varepsilon(t)$, $m_D(t)$,  and $\tstar(t)$ as functions of time rescaled  by the  thermalization timescale  $\tauT =  \nicefrac{\alpha_s^{\nicefrac{-13}{5}}}{\Q_s}$ \cite{Baier:2000sb,boguslavski2023limiting}. The results are shown in \fig \ref{fig:comparison}. The main result is that for the same $\varepsilon$ (Landau matching), the deviation from equilibrium  is $\sim 30 \%$ (left panel). When matching for the same $m_D$ (center panel) or $\tstar$ (right panel) the deviations are considerably larger.
\def\figsize{0.26}
\begin{figure}
\includegraphics[scale=\figsize]{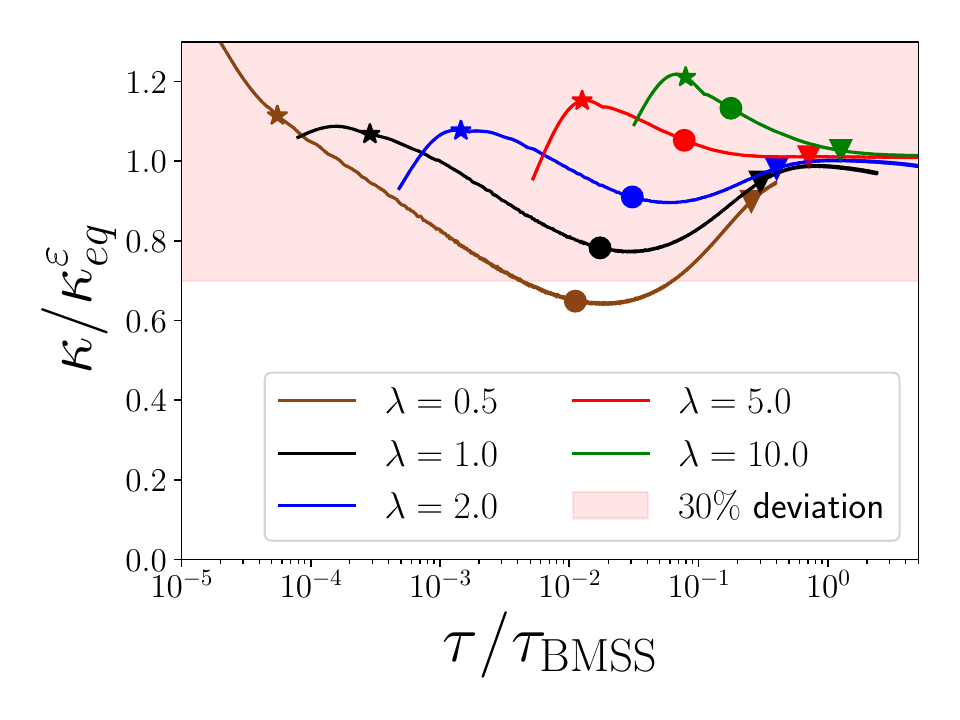}
\includegraphics[scale=\figsize]{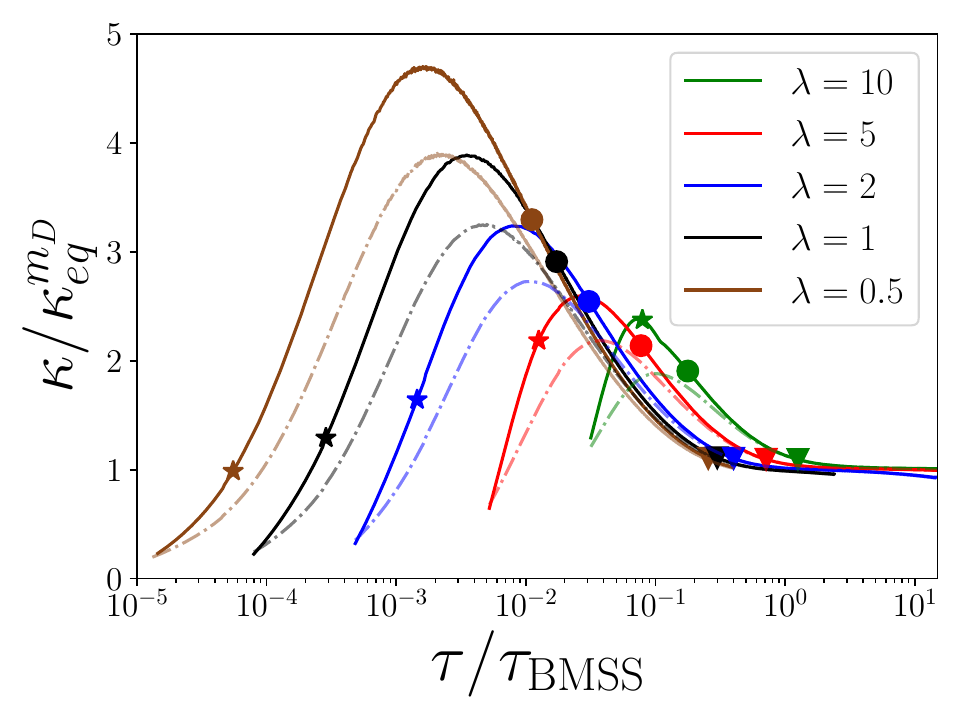} 
\includegraphics[scale=\figsize]{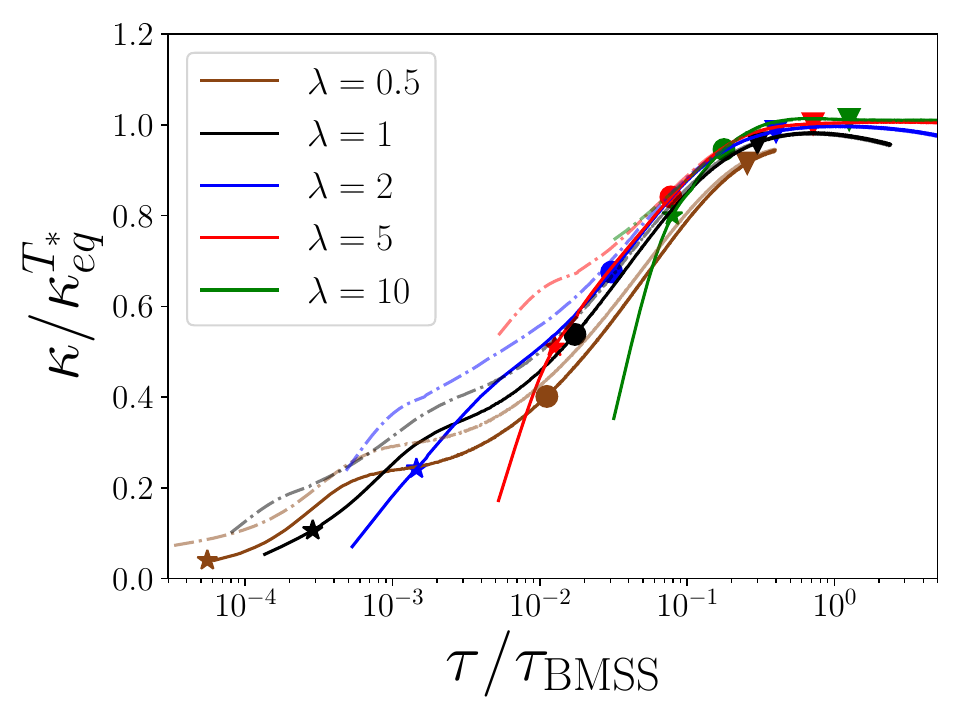}
\caption{Equilibrium and nonequilibrium $\kappa$ for the same $T_\varepsilon$ (left),  $m_D$ (center) and $\tstar$ (right). Figures taken from \cite{Boguslavski:2023fdm}. }
\label{fig:comparison}
\end{figure}

\subsection{Transverse vs. longitudinal diffusion coefficient}
The ratio of the transverse and longitudinal diffusion coefficients is shown in the left panel in \fig \ref{fig:moneyplots}. The initial $\kappa_T/\kappa_z > 1$ arises from the overoccupation  and large anisotropy leading to enhanced transverse momentum exchange. After the star marker, i.e., during the second stage of the bottom-up scenario, one finds   $\kappa_T/\kappa_z < 1$. This originates from the large momentum anisotropy of the underoccupied system and is in line with results from squeezed thermal distributions  \cite{Romatschke:2006bb}. Between a maximal underoccupation and hydrodynamization, the ratio smoothly evolves towards unity. Throughout the whole evolution, the anisotropy ratio is at most 2.

\subsection{Comparison with lattice \& glasma}
\begin{figure}
\includegraphics[scale=\figsize]{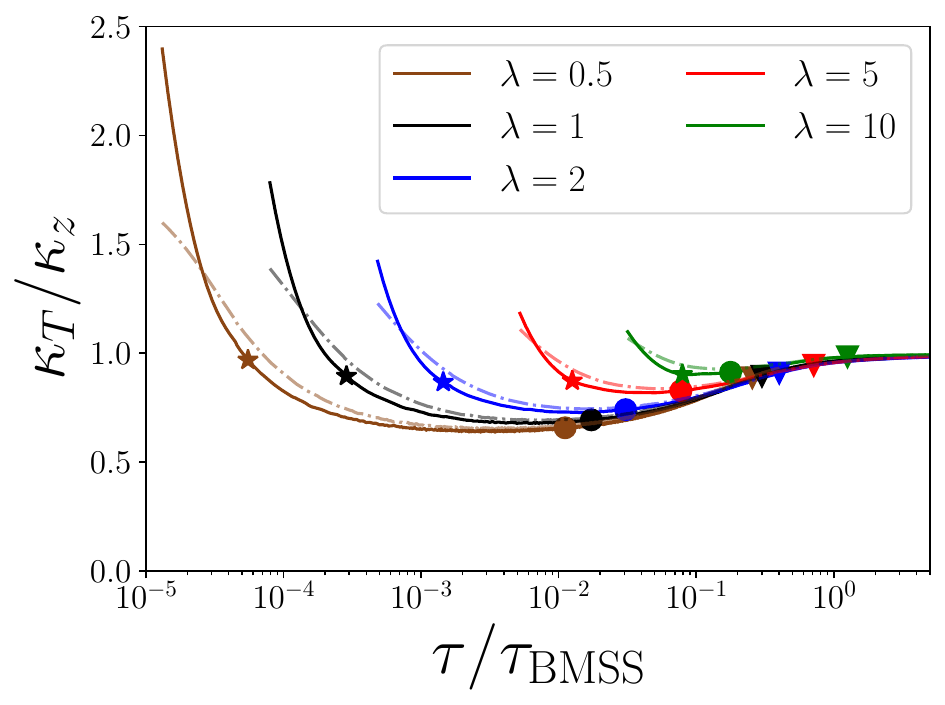}
\includegraphics[scale=\figsize]{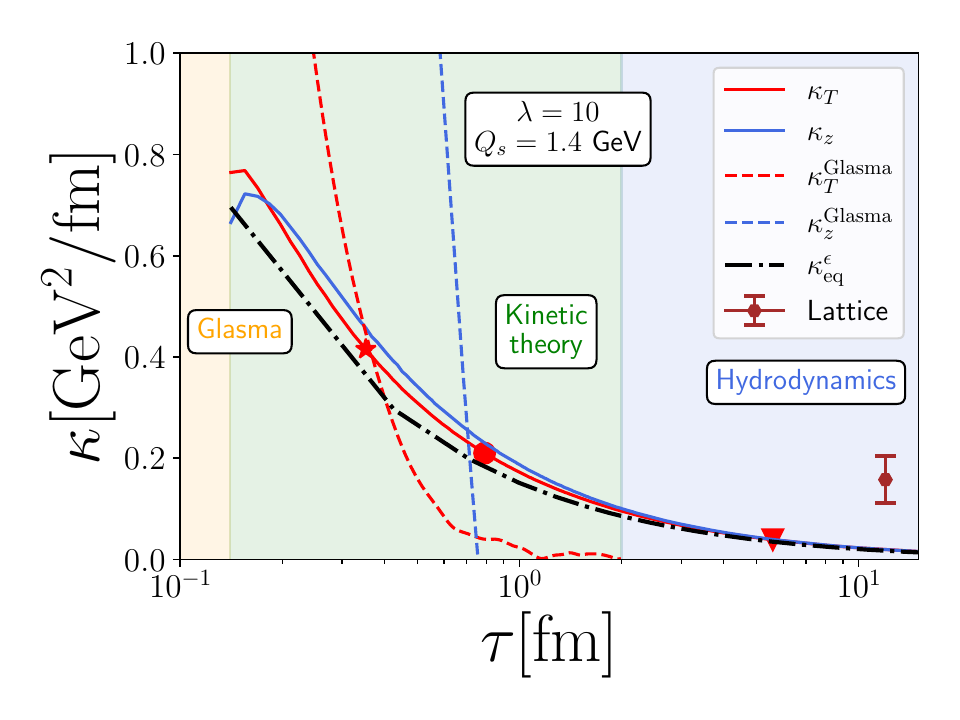}
\includegraphics[scale=\figsize]{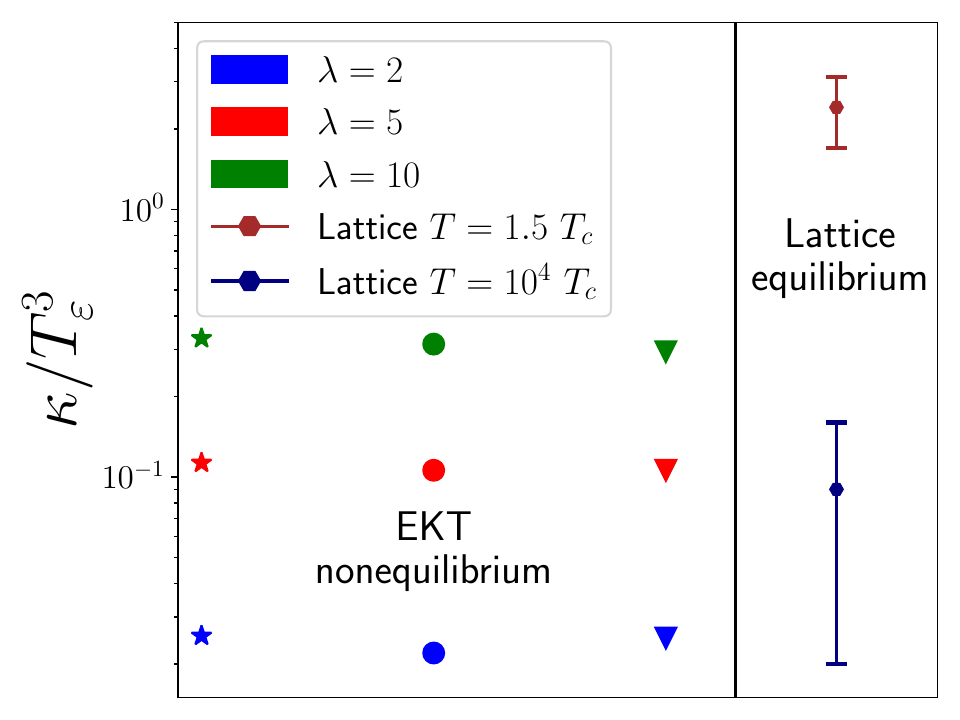}
\caption{Left: Comparison of transverse and longitudinal diffusion coefficients as functions of time in the units of the thermalization time. Center: Comparison of our results and \cite{Avramescu:2023vld} in units of $\mathrm{GeV}$.  Right: Comparison of our values with lattice simulation results \cite{Brambilla:2022xbd}.   Figures taken from \cite{Boguslavski:2023fdm}.  }
\label{fig:moneyplots}
\end{figure}
The center panel of \fig \ref{fig:moneyplots} shows our and the glasma  results \cite{Avramescu:2023vld}, which are initially considerably larger. The transverse diffusion coefficients  match better than the longitudinal ones. The lattice result \cite{Brambilla:2022xbd} at $T = 1.5T_c$ is depicted for the same energy density.

The right panel of \fig \ref{fig:moneyplots} shows our results for  $\lambda=2,5,10$ and lattice results \cite{Brambilla:2022xbd} at $T= 1.5T_c$ and $T= 10^4T_c$ in terms of the ratio $\nicefrac{\kappa}{T^3}$. At extremely high temperatures, the coupling of the lattice calculation  corresponds to $\lambda \approx 2$, while for the lower temperature $\lambda \sim 10$ from the one-loop beta function (breaks down at this scale). 
The stages of the bottom-up evolution are shown by the respective markers. Our result for  $\lambda = 2$ 
is in rough agreement with the lattice estimate at $T= 10^4T_c$. However, at $1.5T_c$ the lattice result is considerably larger.

\section{ Conclusions \& outlook}
Our primary aim in this paper is to understand the magnitude and anisotropy of $\kappa$ during hydrodynamization. We find that the diffusion coefficient is within 30 \% from its equilibrium value for the same energy density. For the anisotropy of the diffusion coefficient, we observed that initially $\kappa_T > \kappa_z$. For underoccupied systems, the hierarchy is reversed. The maximal difference between $\kappa_T$ and $\kappa_z$ throughout the entire evolution is a factor of $\lesssim 2$.

We expect our results to have applications especially in phenomenological descriptions of heavy quark diffusion and quarkonium dynamics. 

\section*{Acknowledgements}
This work is supported by the European Research Council, ERC-2018-ADG-835105 YoctoLHC and under the European Union’s Horizon 2020 research and innovation by the STRONG-2020 project (grant agreement No. 824093), Academy of Finland by the Centre of Excellence in Quark Matter (project 346324) and project 321840, the Austrian Science Fund (FWF) under project P 34455, and the Doctoral Program W1252-N27 Particles and Interactions.
The authors wish to acknowledge CSC – IT Center for Science, Finland, for computational resources. The content of this article does not reflect the official opinion of the European Union and responsibility for the information and views expressed therein lies entirely with the authors.

\bibliography{QMBIB}

%

\end{document}